\documentclass[
    ,final            
  ]
  {aipproc}

\layoutstyle{6x9}

\begin{document}

\title{Multicolour photometry of the M0V+M5V eclipsing binary V405 And}


\author{K. Vida}{
  address={E\"otv\"os Lor\'and University, Department of Astronomy,  Budapest,Hungary},
  altaddress={Konkoly Observatory, Budapest, Hungary}
}

\author{K. Ol\'ah}{
  address={Konkoly Observatory, Budapest, Hungary}
}

\author{Zs. K\H{o}v\'ari}{
  address={Konkoly Observatory, Budapest, Hungary}
}

\author{J. Bartus}{
  address={Astrophysikalisches Institut Potsdam, Germany}
}

\begin{abstract}
 We present multicolour photometry and modelling of the active eclipsing binary star V405 And. The components of 0.2 and 0.5 solar masses are just below and above the theoretical limit of the full convection, that is thought to be around 0.3 solar mass. The light curves are compositions of constant and variable features: the distorted shape of the components (about 25\%), a small eclipse, and mainly of spots (about 75\%) and flares.
\end{abstract}

\maketitle


\section{Introduction}

The manifestations of stellar activity appearing as spots, plages, flares, activity
cycles etc., are consequences of the strong magnetic fields. The origin of this magnetic field of
stars is some kind of a dynamo mechanism, which can be the so-called $\alpha\Omega$ dynamo, based
on the amplification of the fields by differential rotation in the tachocline, which is a
thin interface layer between the convection zone and
the radiative core. Less massive stars are thought to be fully convective, wherein the $\alpha^2$ dynamo
generates strong, long-lasting, axisymmetric magnetic fields. 
The mass limit for full convection is thought to be around ~0.35 solar mass.
The late type binary V405~And is an excellent target to study the effects of the two possible
dynamos, since one component is below, the other is above the mass limit of the full convection \cite{1997A&A...326..228C}.	

\section{Observations and data reduction}

\begin{figure}
 	\centering
 \includegraphics[width=\textwidth]{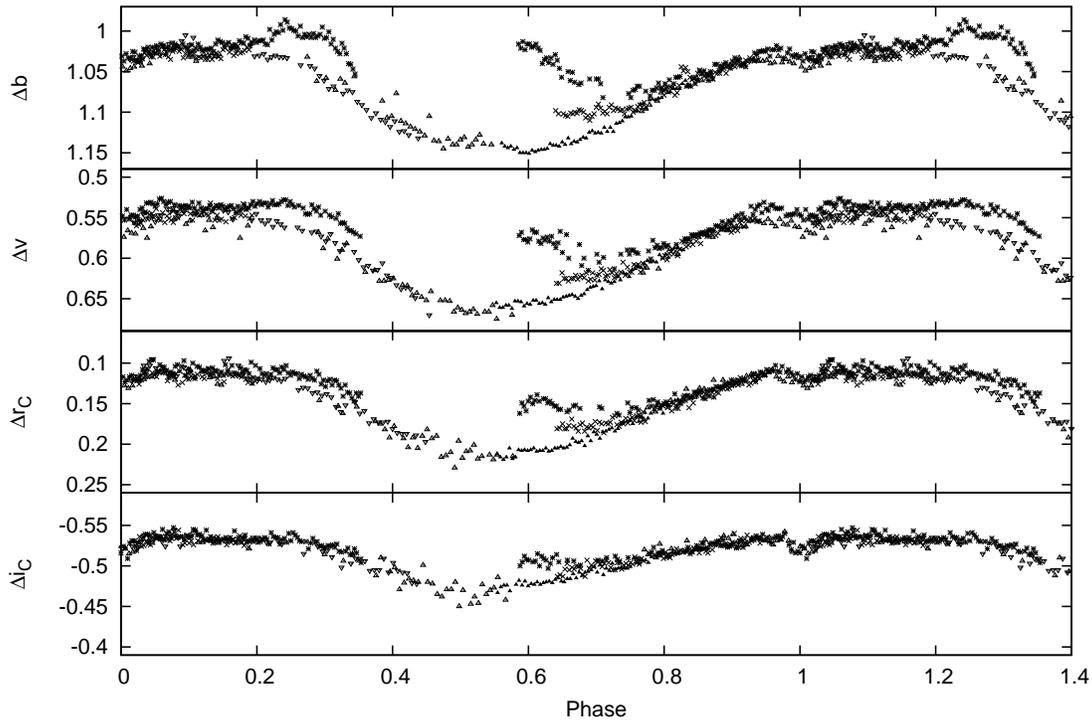} 
 \caption{Observational data obtained with the 1m RCC telescope in Konkoly Observatory. Each
observed night was plotted with different symbols. Note the small eclipse at phase 1.0. The light
curve show the slow flare observed on JD 2454373 and JD 2454374 starting at phase $\approx$0.6. (see
Fig. \ref{fig:flares}b)}
 \label{fig:observations}
\end{figure}

Observations were carried out using the 1m RCC-telescope of the Konkoly Observatory at Piszk\'estet\H{o} Mountain Station. The telescope is equipped with a Princeton Instruments $1300\times1300$ CCD. We present data obtained on $5$ nights (between JD 2454373--2454426). We also present one night of observation (at JD 2454141) obtained with the 60 cm telescope of the Konkoly Observatory, Sv\'abhegy, Budapest equipped with a Wright Instruments
$750\times1100$ CCD camera.
For differential photometry GSC 03298-00148 was used as comparison, GSC 03298-01110 as check star. 
Data reduction was done using standard IRAF\footnote{
IRAF is distributed by the National Optical Astronomy Observatory, which
is operated by the Association of Universities for Research in Astronomy, Inc.,
under cooperative agreement with the National Science Foundation.
} packages and DAOPHOT was used for photometry. The
resulting light curve is plotted in Fig. \ref{fig:observations}.

\section{Analysis of the data}
For modelling the data, we used the light curve between HJD 245422--245426 from the RCC
telescope, free from flare activity. From Fig. 5 in \cite{1997A&A...326..228C} it is seen that the light curve of V405~And is
quite stable, and since our data used for eclipse modelling is observed within 5 days, the effect
of spot changes, thus the modulation of the light curve is probably small. Since the light curve
is composed of eclipse/shape distortion and of modulation caused by starspots,
parallel modelling of both effects is needed \cite{2008A&A...485..233S}. Basic binary
parameters (semi-major axis, mass ratio, orbital period) were known from the radial velocity curve
\cite{1997A&A...326..228C}, thus we had good assumptions on stellar radii and system inclination.
Temperatures were estimated from the spectral types using the tables of \cite{1996ApJ...469..355F} and \cite{2003AJ....126..778V}.
The
binary solution was carried out using PHOEBE \cite{2005ApJ...628..426P}. The derived system
parameters are summarised in Table \ref{tab:params}. A spot model has been
calculated using the program SML (SpotModeL, \cite{2003AN....324..202R}) assuming the presence of two cool circular spots with $T_{spot}=3300$K, on the primary star. This assumption is plausible, since the spottedness of the smaller, fainter companion wouldn't affect the shape of the light curve much.

The light curve solution of the was carried out in the following way: first, a binary model is substracted from the observed light curve. Since the resulting light
curve is free from the effects of the binarity -- the eclipse and the distorted shape, the only cause
of the changes is the spottedness of the primary star.  In the next step a spot model is applied,
which is then removed from the original, observed light curve. The result is an unspotted
binary light curve, for which a more precise binary model can be derived.  After some iterations
times the light curve can be fitted with a composite solution of the spots and binarity. 
During the procedure the unspotted brightness of the system should be taken into account, which, in our case is the brightest magnitude ever observed \cite{1997A&A...326..228C}.

\begin{figure}
 \centering
 \includegraphics[width=0.5\textwidth]{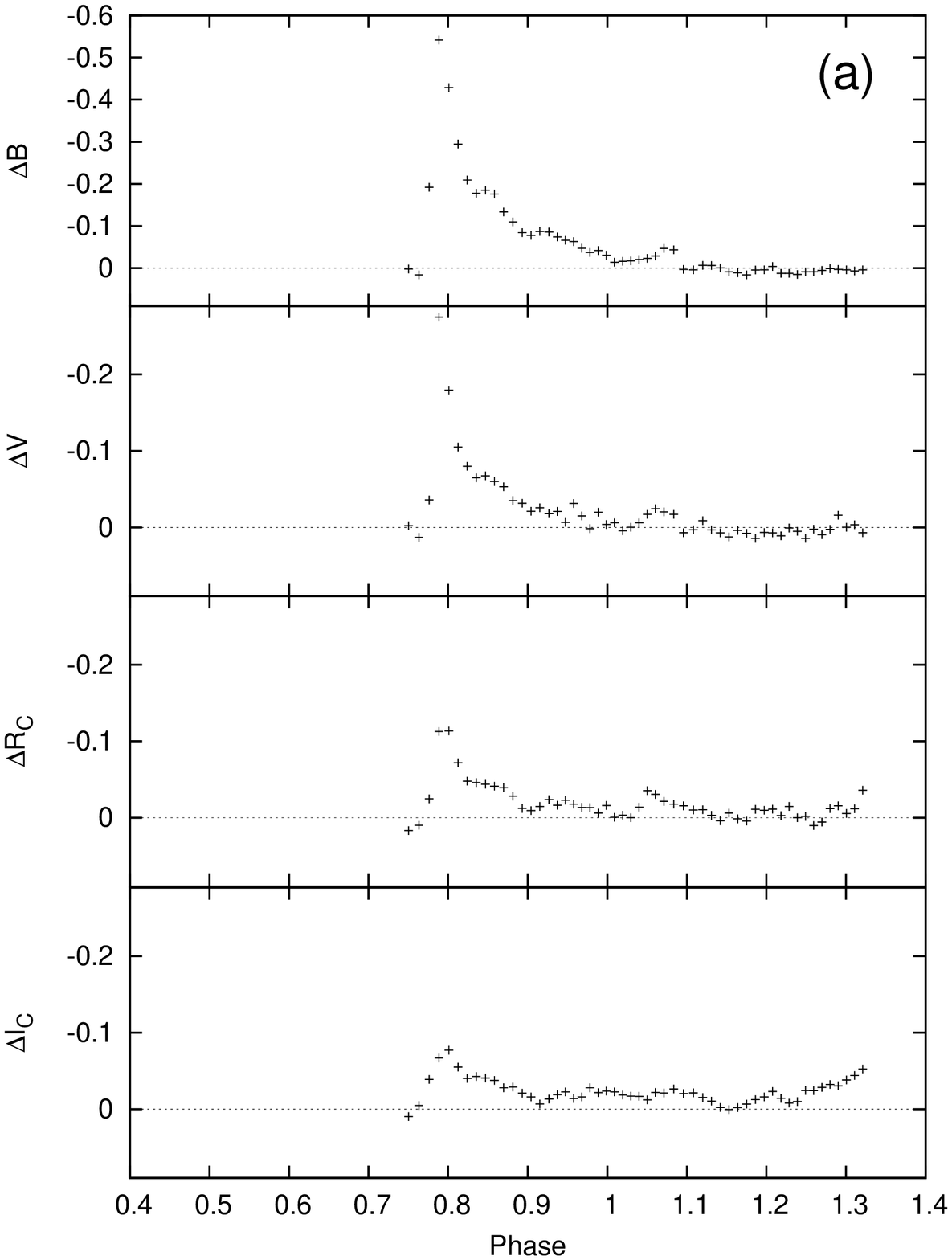}
 \includegraphics[width=0.5\textwidth]{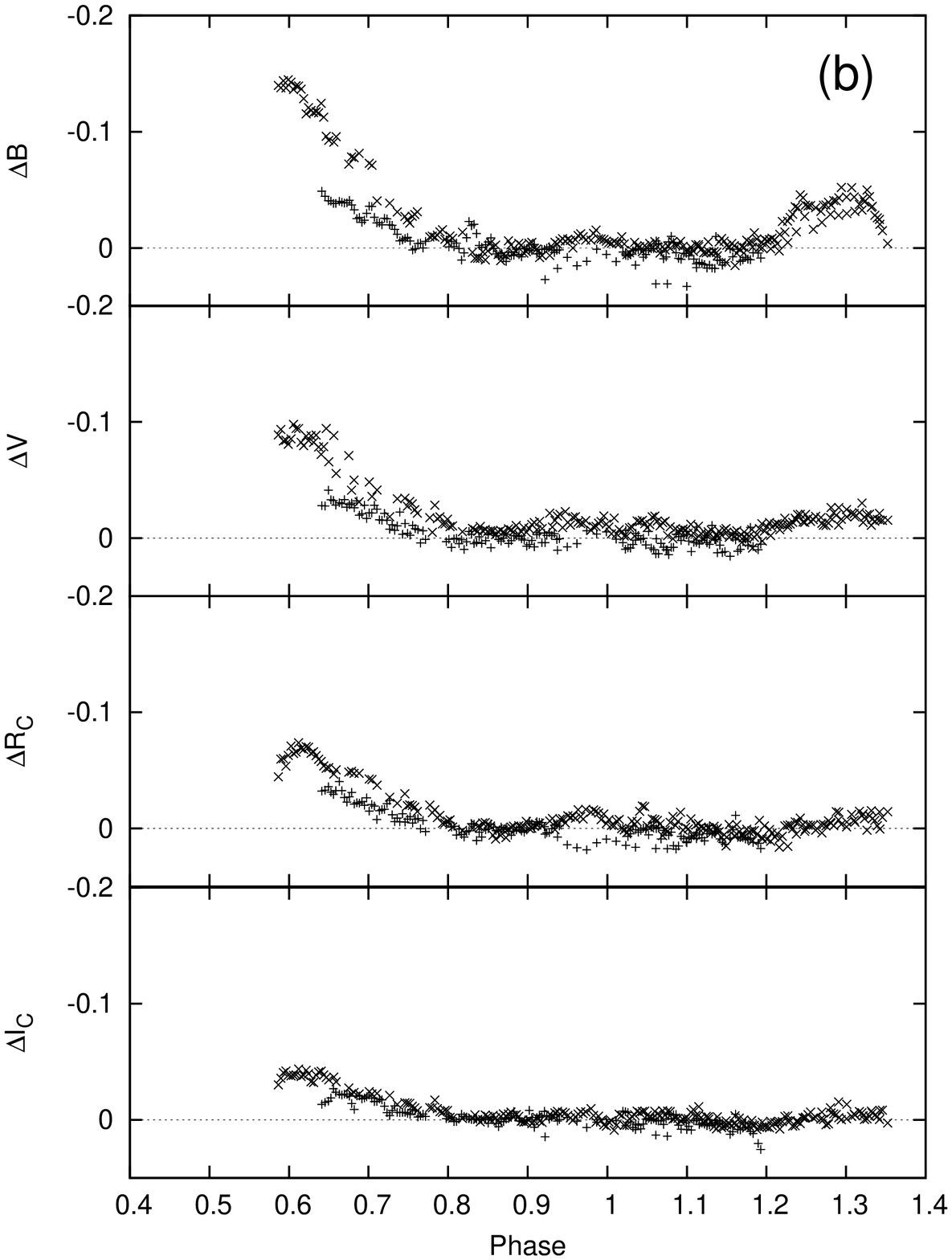}
 \caption{(\textbf{a}) A giant flare observed at HJD
2454141 using the 60cm telescope of Konkoly Observatory. (\textbf{b}) A long lasting (at
least 3 days) slow flare event from the 1m RCC data observed at HJD 2454373 (plus signs)
and at HJD 2454374 (crosses). 
Binary+spot model light curves are subtracted from the original data, a line
represents the undisturbed stellar surface.}
 \label{fig:flares}
\end{figure}

In the observations, two remarkable flare events occurred. The light curves of the flares are plotted in Fig. \ref{fig:flares}, where the binary and spot models are substracted. The estimated flare energies for the larger one appeared at phase $\approx0.8$ (Fig. \ref{fig:flares}a) in $B,V,R_C$ and $I_C$ passbands are $14.10\times 10^{35}, 7.38\times 10^{35}, 13.33\times 10^{35}$ and $13.25\times 10^{35}$ergs/sec, respectively (for the description of the estimation method, see e.g. \cite{2007AN....328..904K}). 

In the case of the second, much slower flare event, starting at phase $\approx0.6$ (Fig. \ref{fig:flares}b) the phenomenon remains observable for at least 3 rotations.
During this time, the flaring region moves out of view at phase $\approx0.9$.

\begin{table}
\centering
\begin{tabular}{rll}
\hline
$P_{orb}=P_{rot}$ &=& $0.465$ days\tablenote{taken from \cite{1997A&A...326..228C}} \\
a &=& 2.246$R_\odot$  $^*$\\
$i$&=& 66.5\\
$M_1$ &=& $0.495 M_\odot$\\
$M_2$ &=& $0.208 M_\odot$\\
$T_1$ &=& $4050$ K\\
$T_2$ &=& $3000$ K\\
$R_1$ &=& $0.785 R_\odot$\\
$R_2$ &=& $0.235 R_\odot$\\
$v\sin i$ &=& 85 km/s\\
$e$ &=& $0.0^*$ \\
\noalign{\smallskip}
\label{tab:params}

\end{tabular}
\caption{System parameters for V405 And}
\end{table}


\begin{theacknowledgments}
KV, KO and are supported by the Hungarian Science Research Program (OTKA) grant T-048961.
ZsK is a grantee of the Bolyai J\'anos Scholarship of the Hungarian Academy of Sciences.
\end{theacknowledgments}

\bibliographystyle{aipproc}   

\bibliography{mn-jour,mybib}

\end{document}